\documentclass[prb,twocolumn,aps,superscriptaddress,showpacs,floatfix]{revtex4}
\usepackage[centertags]{amsmath}
\usepackage{amssymb}
\usepackage{amsthm}
\usepackage{graphicx,subfigure}
\usepackage{epsfig}

\begin{document}

\title{Periodic force induced stabilization or destabilization of the denatured state of a protein}

\author{Pulak Kumar Ghosh$^a${\footnote{e-mail:gpulakchem@gmail.com}},
  Mai Suan Li$^c${\footnote{e-mail: masli@ifpan.edu.pl} }
  and
Bidhan Chandra Bag$^{b}$ {\footnote{e-mail:
pcbcb@rediffmail.com}}}

 \affiliation{
$^a$Advanced Science Institute, RIKEN, Wako, Saitama, 351-0198, Japan\\
$^b$ Department of Chemistry, Visva-Bharati, Santiniketan 731 235,
India\\
$^c$Institute of Physics, Polish Academy of Science, Poland}

\begin{abstract}
We have studied the effects of an external sinusoidal force in
protein folding kinetics. The externally applied force field acts
on the each amino acid residues of polypeptide chains. Our
simulation results show that mean protein folding time first
increases with driving frequency and then decreases passing
through a maximum. With further increase of the driving frequency
the mean folding time starts increasing as the noise-induced
hoping event (from the denatured state to the native state) begins
to experience many oscillations over the mean barrier crossing
time period. Thus unlike one-dimensional barrier crossing
problems, the external oscillating force field induces both
\emph{stabilization or destabilization of the denatured state} of
a protein. We have also studied the parametric dependence of the
folding dynamics on temperature, viscosity, non-Markovian
character of bath  in presence of the external field.
\end{abstract}

 \maketitle
\section{Introduction}
Enhancement of reaction kinetics due to interplay between barrier
fluctuation rate and thermal noise-assisted barrier crossing
events known as resonant activation \cite{dor}, is an interesting
observation in early 1990s. This phenomenon provides a better
understanding the mechanisms of various chemical and biological
processes. Examples include: dissociation kinetics of large
molecules in coupled chemical systems\cite{Maddox}, oxygen binding
mechanisms to hemoglobin\cite{Beece}, modelling the dynamics of
dye laser,
 ratchet models for the directional movements of molecular motors,
transport through artificial nanopores \cite{Marchi0} etc. The
interesting mechanism for enhancing rate via resonant activation
has triggered a numerous theoretical investigations for  Markovian
and non-Markovian nature of the bath and external deriving forces
\cite{bier,van,Marchi,brey,klaf,pk1}. This phenomenon has been
experimentally realized \cite{exp1} in a tunnel diode biased in a
strongly asymmetric bistable state in the presence of two
independent sources of electronic noise. The interference effects
of resonant activation and stochastic resonance have also been
studied to the aim for stochastic localization of particles
confined in a bistable potential as well as in a multi-well
system. \cite{loca,loca1}.

The overwhelm majority of the previous studies on barrier crossing
dynamics over a fluctuating energy barrier are based on the
dynamics of a molecule having a single degree of freedom. It is
modeled by a Brownain particle in a bistable or metatable
potential with fluctuating barrier. But dynamics of the molecules
having large number of degrees of freedom offers a significantly
different situation, due to its structural rigidity and several
interaction energies. Here, in addition to the energetic barrier
and fluctuation statistics, the dynamics of the molecules is
largely controlled by entropic factors.

The folding dynamics of macromolecules like proteins is an example
of thermally activated barrier crossing dynamics in a
multidimensional space. Here, all the degrees of freedom are
coupled to each other through various interactions, such as,
nearest-neighbor interactions,  dihedral potentials, hydrogen
bonds, ion pairs, van der Waals interactions etc. Over the years a
considerable attention has been focused on better understanding of
folding mechanisms and potential energy landscapes. The studies of
protein folding dynamics under different external conditions such
as salt concentration, temperature, confinement, pH and viscosity
of the medium etc
 \cite{onu,sha,bry,thir,li1,cie,li2,kli1,kli2,vel,Clementi} provide
important information about its structure and functionality.
Again, the single molecule pulling experiments by highly sensitive
force probes such as atomic force microscopy\cite{florin,chen,kumar} and
optical and magnetic tweezers\cite{mehta,danil} make it possible
to realize various interactions due to internal degrees of freedom
of a macromolecule. Motivated by the recent
studies\cite{florin,chen,mehta,danil} in this context we have
investigated the effect of an applied external periodic field of
non-thermal origin on the protein folding dynamics. We have
assumed that the external electric field interacts with charge or
dipole moment of the amino acid residues.

To accomplish our goal we have considered the well-known
coarse-grained off-lattice models\cite{hon,Clementi} for
polypeptide chain (where each amino acid residue is considered as
a single bead centered at their $C^{\alpha}$ position). The
interaction energies which play roles in the folding dynamics are
taken into account by a Go-like Hamiltonian\cite{go}. By setting
the Langevin equations for each bead we have followed the dynamics
of protein chains in presence of external fluctuations (field).

A number of experimental studies\cite{Plaxco_PNAS98,Pradeep_JMB07}
imply that Markovian dynamics cannot accurately account the effect
of viscosity on the barrier crossing phenomenon in a solution
phase and the theory based on the non-Markovian dynamics shows a
better agreement between theoretical and experimental results.
Therefore, to capture the important effects of non-Markovian
dynamics and also for a sake of generality we have considered
exponentially decaying memory of the thermal fluctuations.
Moreover, the present study is an extension of earlier studies
\cite{astumiam1,astumiam2,bagphysA,zolo} on barrier crossing
dynamics in low dimensional systems in presence of periodic force.
To compare the features of folding kinetics of macromolecules with
one dimensional barrier crossing problems, we also simulate a
non-Markovian Langevin dynamics in a bistable potential in
presence of barrier fluctuations.

Specifically our objective here is twofold. First we intend to
explore the effects of an applied external oscillating field on
protein folding to the goal of extracting generic features of
multidimensional barrier crossing dynamics in contrast to the one
dimensional cases. The second objective is to investigate the
effects of the oscillating time periodic force field on the
following three important features of protein folding kinetics:
(i) turnover behavior of mean folding time with solvent viscosity,
(ii) U-shaped mean folding time \emph{vs.} temperature profiles,
and (iii) double minimum in mean folding time as a function of the
correction of thermal fluctuations.

To address the above challenging issues we simulate folding
dynamics of the 16-residue peptide $\beta$-hairpin (C-terminal
from protein G, PDB ID: 2gb1) and 76-residue protein ubiquitin
(PDB ID: 1ubq). The PDB structures of these proteins are shown in
Fig.~1.

\begin{figure}[!htb]
\includegraphics[width = 8cm,angle=0,clip]{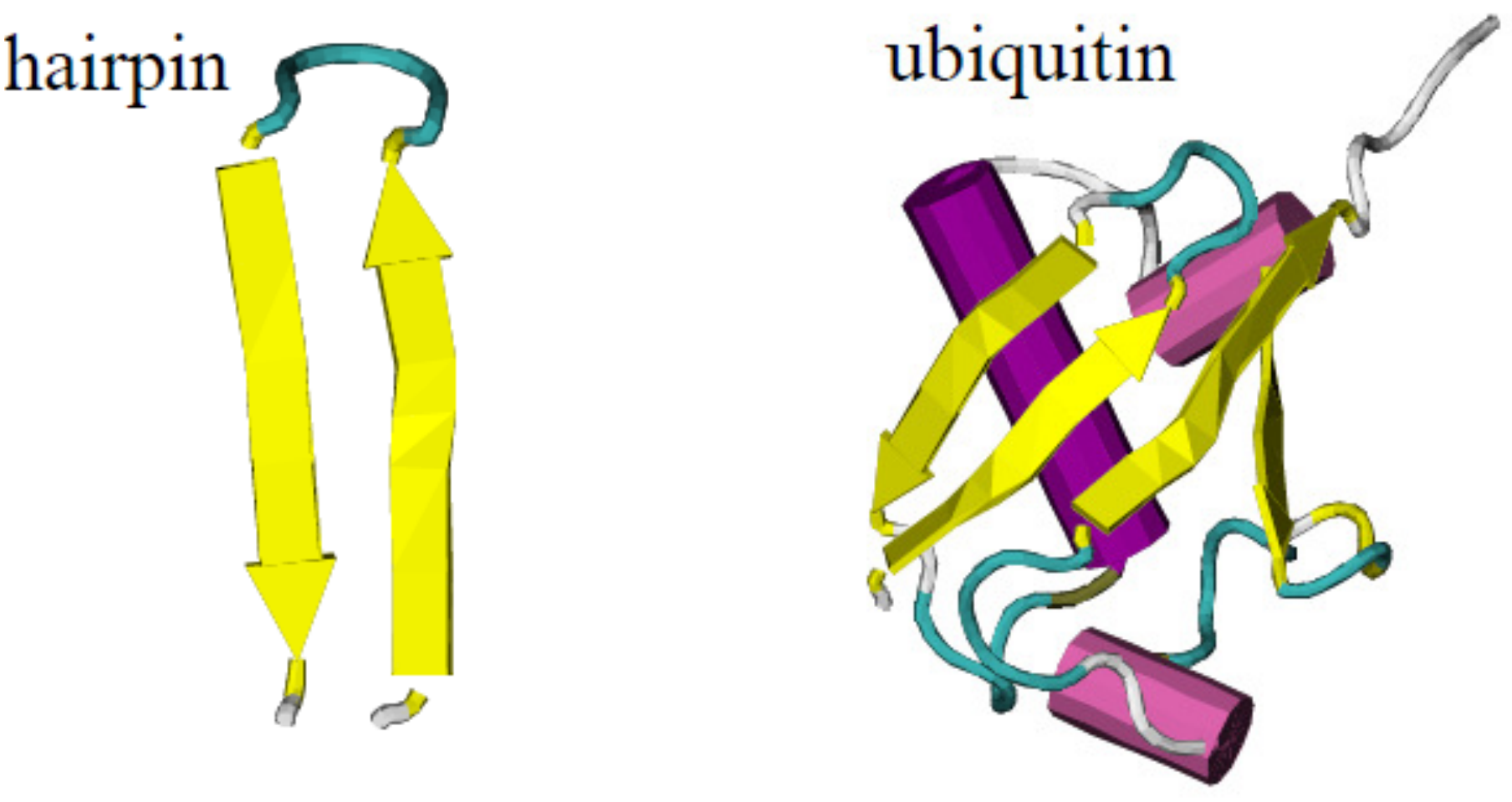} \caption{
(Color online) The PDB structures of two proteins studied in this
work. For the cutoff distance $d_c\; =\; 6.5 \;\AA$ , the total
number of native contact is equal $Q_{max}=13$ and $99$ for
hairpin and ubiquitin , respectively. } \label{fig0}
\end{figure}

\section{The Model}
To study protein folding dynamics we consider coarse-grained
off-lattice models\cite{Clementi} for polypeptide chains in which each
amino acid residue is represented as a single bead centered at its
$C^{\alpha}$ position. Moreover, we follow the dynamics of
polypeptide chains  by Go-like Hamiltonian\cite{go}, in which the
interactions between residues forming native contacts are assumed
to be attractive and non-native interactions are repulsive. The
energy of a configuration of a protein is specified by the
coordinates $r_i$ of the $C^{\alpha}$ atoms and is given by \cite{Clementi}
\begin{eqnarray}
E&=&\sum_{bonds}K_r\left( r_{i,i+1}-r_{0i,i+1}\right)^2\;+
\;\sum_{angles}K_{\theta}\left(\theta_i-\theta_{0i}\right)^2\nonumber
\\
&+& \sum_{dihedral}K_{\phi}^{(1)}\left[ 1-\cos{
(\phi_i-\phi_{0i})}\right]\; + \sum_{dihedral}\;K_{\phi}^{(3)}
\nonumber\\
&\times&\left[ 1-\cos{(
3(\phi_i-\phi_{0i}))}\right]\;+\sum_{i<j-3}^{NC}\epsilon_H\left[
5R_{ij}^{12}- 6R_{ij}^{10}\right]
\nonumber\\
&&
+\sum_{i<j-3}^{NNC}\epsilon_H\left(\frac{C}{r_{ij}}\right)^{12}\label{1.1}
\end{eqnarray}

$r_{i,i+1}$ is the distance between $i$-th and $(i+1)$-th beads.
$\theta_i$ is the bond angle formed by three subsequent beads:
$(i-1)$-th, $i$-th and $(i+1)$-th.  $\phi_i$ denotes the dihedral
angle around the $i$-th bond and $r_{ij}$ is the distance between
the $i$-th and $j$-th residues. The subscripts $0$, $NC$, and
$NNC$ refer to the native configuration, the native contact, and
the non-native contact, respectively. $r_{0ij}$ is the distance
between the $i$-th and  the $j$-th  residues in the native
conformation and $R_{ij}=\frac{r_{0ij}}{r_{ij}}$. Amino acid
residues are assumed to be in the native contact if $r_{0ij}$ is
less than a given cutoff distance$(d_c)$. Here, we have assumed
the cutoff distance $d_c=6{\AA}$.

The first term of Eq.(\ref{1.1}) presents harmonic potential due
to chain connectivity between two adjust beads. The second term is
also harmonic potential arising due to bond angle between three
subsequent beads. The third term presents dihedral potential for
every four adjacent $C^{\alpha}$ atoms. Dihedral potential is a
sum of two periodic components with periods:
$\tau_{\phi}=2\pi/(\phi_i-\phi_{0i})$ and
$\tau_{\phi}=2\pi/3(\phi_i-\phi_{0i})$. The last two terms in
Eq.(\ref{1.1}) are due to the non-local native interactions and
the short-range repulsive force for non-native pairs,
respectively. Parameters
$K_r,\;K_{\theta},\;K_{\phi},\;\epsilon_H$ denote the relative
strength of each kind of interaction; we choose
$K_r=100\epsilon_H/{\AA}^2,\;K_{\theta}=20\epsilon_H/rad^2,\;K_{\phi}^{(1)}=\epsilon_H
,\;K_{\phi}^{(3)}=0.5\epsilon_H$, where $\epsilon_H$ is the
characteristic hydrogen bond energy and $C=4{\AA}$.

The dynamics of the protein chain in a thermal bath can be
described by setting a Langevin equation for each bead (
$C^{\alpha}$-atoms of each amino acid residue) of the protein
chain.
\begin{eqnarray}
m\ddot {\vec{r}}=\vec{F}_c-\int_0^t \gamma(t-t')\dot
{\vec{r}}dt'+\vec{\eta}(t)\label{1.2}
\end{eqnarray}
where $m$ denotes the mass of a bead and $\vec {F}_c=\nabla E$
corresponds to force derived from the potential energy associated
with the protein molecule (Eq.(\ref{1.1})). The potential force
$\vec{F}_c$ depends on the positions of the all  beads. Therefore,
$N$ coupled Langevin equations are needed to follow the time
evolution of a protein having $N$ amino acid residues (beads).
 The thermal fluctuations due to environment
are modelled by colored Gaussian noise. The frictional kernel
$\gamma(t)$ is related to thermal fluctuation $\xi(t)$ by the
well-known fluctuation-dissipation relation.
\begin{eqnarray}
\langle \eta_k(t)\eta_l(t')\rangle=k_B
T\delta_{kl}\gamma(t-t')\label{1.3}
\end{eqnarray}
where, $k_B$ and $T$ are the Boltzmann constant and temperature of
the system, respectively. $\eta_k$ is the $k$-th  components of
$\vec{\eta}(t)$, where $k,l$ = $x,\;y,\;z$; the three orthogonal
components.
 To capture essential
features of the non-Markovian dynamics we consider exponentially
decaying frictional memory
kernel\cite{Bagchi_JCP83,Hansen_book86,Okuyama_JCP86} with the
following form:
\begin{eqnarray}
\gamma(t-t')=\frac{\gamma_0}{\tau}\exp(-\frac{\vert t-t'\vert}{\tau})\label{1.4}
\end{eqnarray}
where $\gamma_0$ is the frictional coefficient in the Markovian
limit and $\tau$ bears the memory effect of the non-Markovian
dynamics.  Then $\vec{\eta(t)}$ is a solution of the following
differential equation: $$\dot{\vec{\eta}}=-\vec{\eta}/\tau+
\frac{\sqrt{\gamma_0 k_B T}}{\tau} \zeta(t).$$ Where, $\zeta(t)$
is a Gaussian white noise having variance two. It should be noted
that for the frictional memory kernel (\ref{1.4}) the
integro-differential (\ref{1.2}) can be simplified as
$m\ddot{\vec{r}}=\vec{F_c} + \vec{\eta}(t)$ and
$\dot{\vec{\eta}}=-\vec{\eta}/\tau-\gamma_0 \dot{\vec{r}}/\tau+
\frac{\sqrt{\gamma_0 k_B T}}{\tau} \zeta(t)$.

We have considered the situation where the dynamics polypeptide
chains is affected by an external sinusoidal force.  It implies
that free energy of the system periodically oscillates in an
asymmetric way. Under such situation the dynamics of the protein
chain is governed by the following equations
\begin{eqnarray}
m\ddot {\vec{r}}=\vec{F}_c-\int_0^t \gamma(t-t')\dot
{\vec{r}}dt'+\vec{\eta}(t)+\vec{A}_0 \sin{\omega t}\label{1.5}
\end{eqnarray}
Unlike the standard pulling experiments (where the external force
is applied to termini of bio-molecules), here the external
oscillatory force filed acts on all amino acid residues. However,
the last term in the above equation accounts effective interaction
between the charge or dipole on the amino acid and electric field.
The resultant force vector corresponding to this interaction
affecting the acceleration vector is parallel to the $\ddot
{\vec{r}}$. Thus $\vec{A}_0$ in the above equation is parallel to
the acceleration vector. We have implemented this in the
simulation scheme.

\subsection{Simulation} In order to analyze the effects of
an external field on the protein folding dynamics we have
calculated the folding time for different parameter sets
(frequency of the external field, dissipation constant of the
medium, temperature, noise correlation time etc.) by numerically
solving the Langevin equation(\ref{1.5}). The relevant
equations(\ref{1.3}) and (\ref{1.5}) were integrated using the
velocity form of Verlet algorithm with the time step $\Delta
t=0.001\tau_L$, $\tau_L$ is the characteristic time scale of the
system which are defined by $\tau_L=(ma^2/\epsilon_H)^{1/2}\approx
3$ ps, where $a$ is the characteristic bond length between two
successive beads $a=4{\AA}$. The mean folding time is the averaged
first passage time from a random configuration (thermodynamically
unstable) to the native state (thermodynamically stable state).
Our calculated mean folding time is the averaged over 250-1000
trajectories depending on the values of different parameters to
have good statistics. To this end we would like mention that in
the present paper we have exploited the numerical scheme of our
earlier study \cite{bagpccp}.

\subsection{Results and discussions}
We have calculated the mean folding time($t_f$)  of
$\beta-$hairpin as a function of frequency of the external field
at temperature $T=0.53 \epsilon_H/k_B=298K$, where
$\epsilon_H=0.98\;K\;cal/mol$ is the hydrogen bond energy. This is
depicted in Fig.2.  The folding time of $\beta-$hairpin first
increases with increasing frequency of the external force
followed by its decrease after passing through a maximum. With
further increase of frequency the folding time starts increasing.
Thus, the mean folding time versus driving frequency profiles
possess both a maximum and a minimum. A similar feature has been
observed (not shown here) at the Markovian limit, $
\tau\rightarrow 0$. It should be noted that the maximum is not
observed in the one-dimensional barrier crossing problem. To
demonstrate this issue and for a qualitative comparison with an
one-dimensional case, we have considered barrier crossing dynamics
of a Brownian particle, modelled by the following generalized
generalized Langevin equation\cite{bagjcp},
\begin{equation}
m\dot{v}=q-q^3-\int_0^t \gamma(t-t')
\dot{v}(t')dt'+\zeta(t)
\end{equation}

\begin{figure}
\epsfig{file=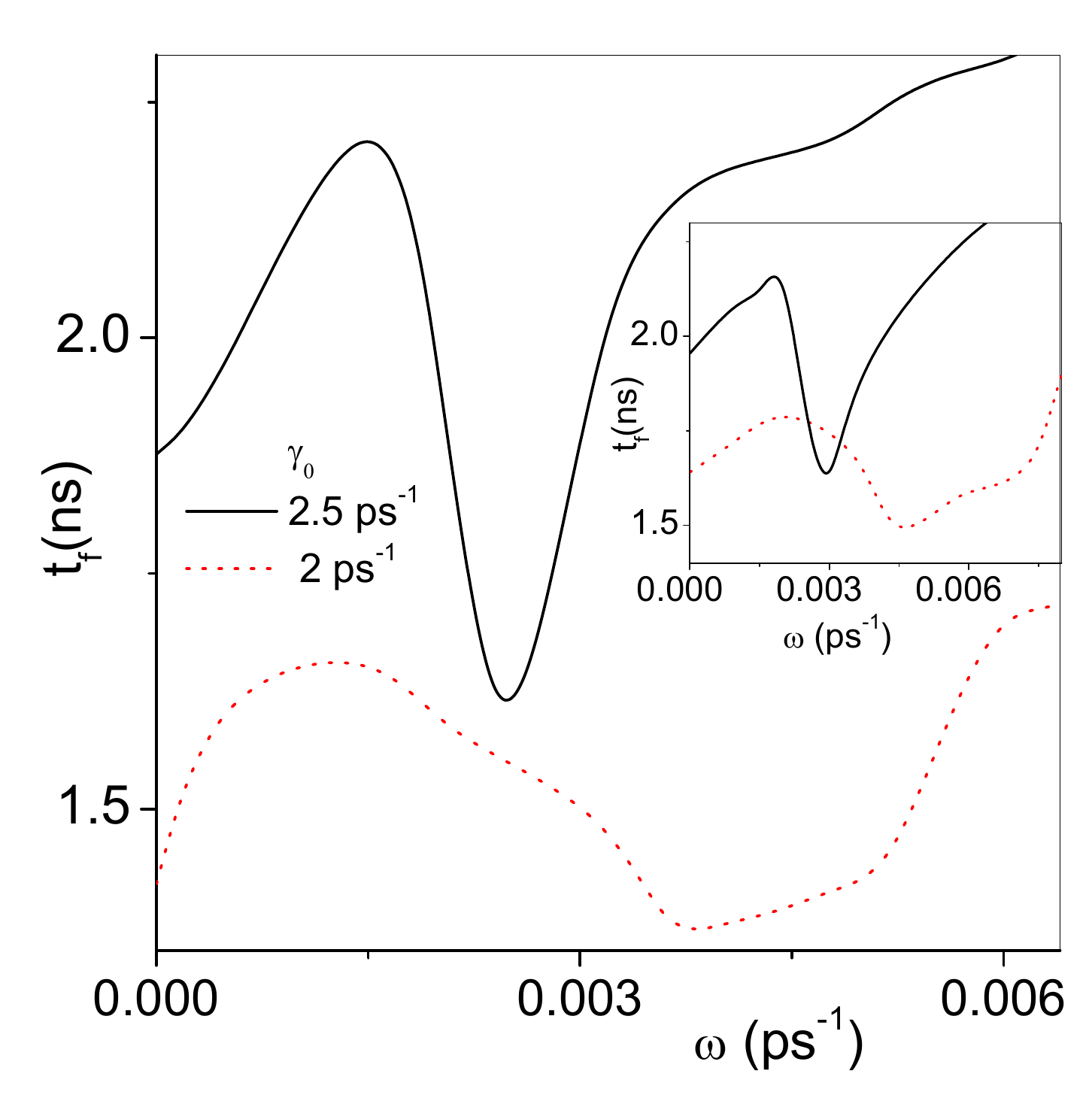,width = 8cm} \caption{(Color online) Plot of
mean folding time $t_f$ of $\beta-$ hairpin vs the frequency of
the external driving force. We chose $T =298K,\; A_0=0.3$
pN,$\;\tau=1.0$ ps. The inset presents the results of Markovian
limit,
 $\tau\rightarrow 0$.} \label{fig1}
\end{figure}
\begin{figure}
\epsfig{file=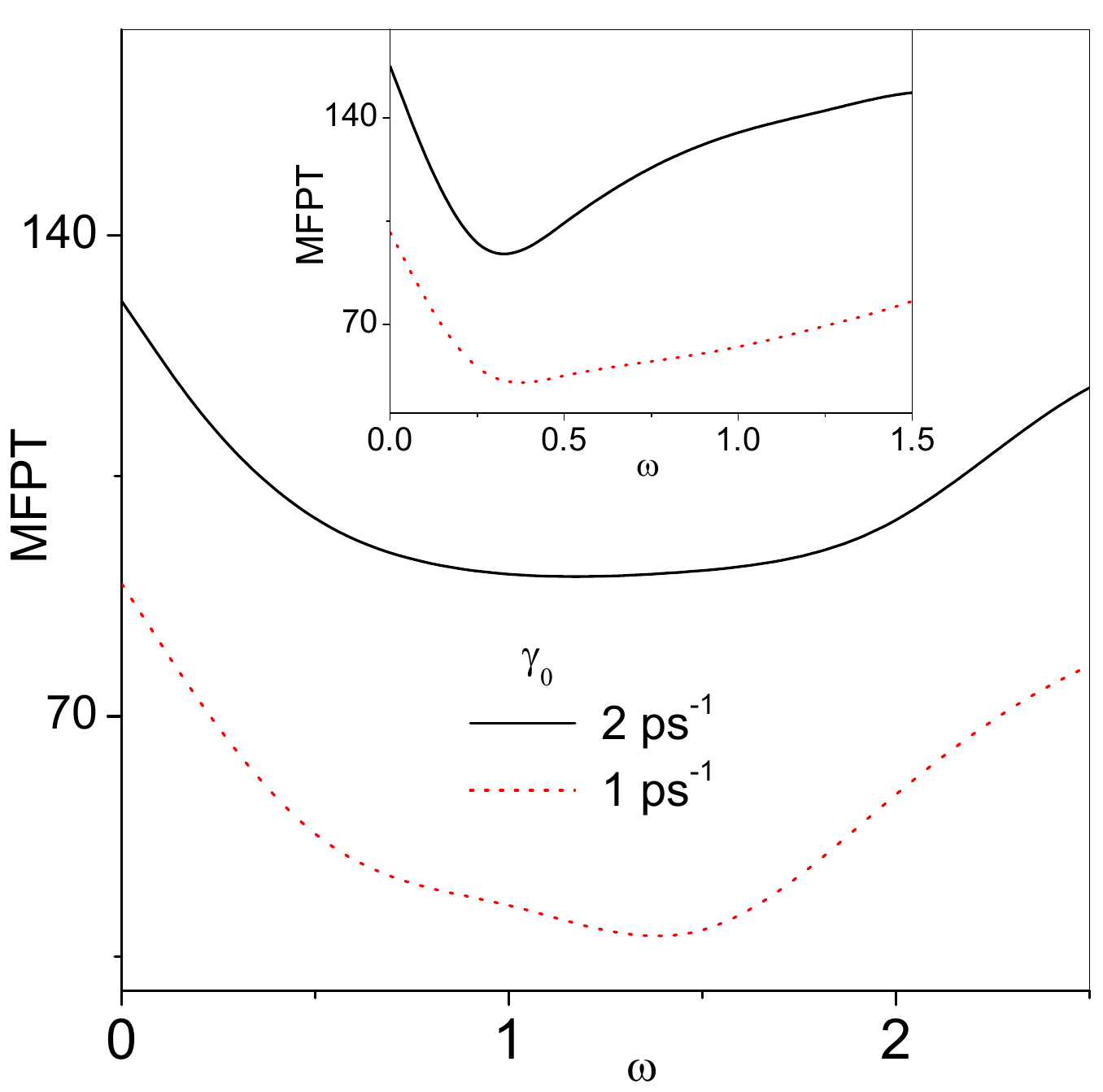,width = 8cm} \caption{(Color online) Plot of
mean first passage time $MFPT$  {\it vs} $\omega$ the frequency of
the external driving force. We chose the double well potential as
$V(q)=1/4q^4-1/2q^2$. We use  $T =0.1,\; A_0=0.25,\;\tau=1.0.$. In
the inset same plot is drawn in the limit $\tau\rightarrow 0$.
(units are arbitrary)} \label{fig2}
\end{figure}

\noindent where $q$ and $v$ are the position and velocity of a
Brownian particle. Here, the folding time is defined as a mean
barrier crossing time, is obtained by solving the above equation
and plotted in Fig.3. The results for the corresponding Markovian
limit have been depicted in the inset of Fig.3. Fig.3 reveals
that, for an one-dimensional case, the mean escape time versus
driving frequency plots possess no maximum. This aspect has also
been reported earlier by several groups using numerical experiment
\cite{bagphysA} and analytical calculation \cite{zolo}. Also, there are
experimental and theoretical studies modelling catalytic reactions
\cite{astumiam2,liu} which show that mean escape time vs. driving
frequency plots exhibit only a minimum but no maximum. Thus,
appearance of the maximum at low frequency regime is a generic
signature of barrier crossing dynamics in many dimensional
systems. In protein folding kinetics, in addition to energetic
barrier, entropy of activation plays a significant role. The
appearance of the maximum (in the $t_f$ \emph{vs.} $\omega$) at
low frequency regime may be attributed to increase of the
fluctuations in configurational entropy during the barrier
crossing rather than activated transport towards the transition
state by the periodic force. For further increase of frequency the
energy transfer to activate towards the transition state dominates
over the increase of fluctuations in configurational entropy and
the mean folding time decreases until the energy transfer rate
becomes small. Thus the maximum and the minimum in $t_f$ vs
$\omega$ plots is a result of interplay between two important
quantities, entropy and activation energy. Similar features of
folding time as a function of driving frequency has been observed
for the higher driving amplitude and damping. It has been
presented in Fig.4.

Figures 2 and 4 reveal that the position of the minimum in the
mean folding time \emph{vs.} frequency plots is very sensitive to
viscosity (damping) of the medium. Positions of the minima are
shifted to the lower frequency with increasing damping. The
pleasurable explanation of this behavior is as follows:  At low
viscosity, because of the strong spatial diffusion the
acceleration of folding kinetics starts to work over the entropy
effect slowly compared to high damping case. Therefore, the
minimum at the higher driving frequency for lower damping constant
is observed.

Our another interesting observation from the Figs.~2 and Fig.~4,
the minimum is shallow and flat for low damping cases compared to
the high damping situation. Because of strong spatial diffusion in
the former case, the effect of the periodic force (by virtue of
doing mechanical work on the particle in the dynamics) is weaken
and it suppresses  activation as well as slow down the variation of
folding time with increase of the driving frequency. Thus, the
increase of the damping constant enhances activation at the cost
of decreasing its robustness.

To generalize our conclusions we  have also studied folding
kinetics of a long protein, ubiquitin. Figure 5 presents mean
folding time of ubiquitin versus driving frequency for different
values of the damping constant. Here also the mean folding time
passes through a maximum and a minimum. Other behaviors are also
very similar to $\beta-$ hairpin.

\begin{figure}
\epsfig{file=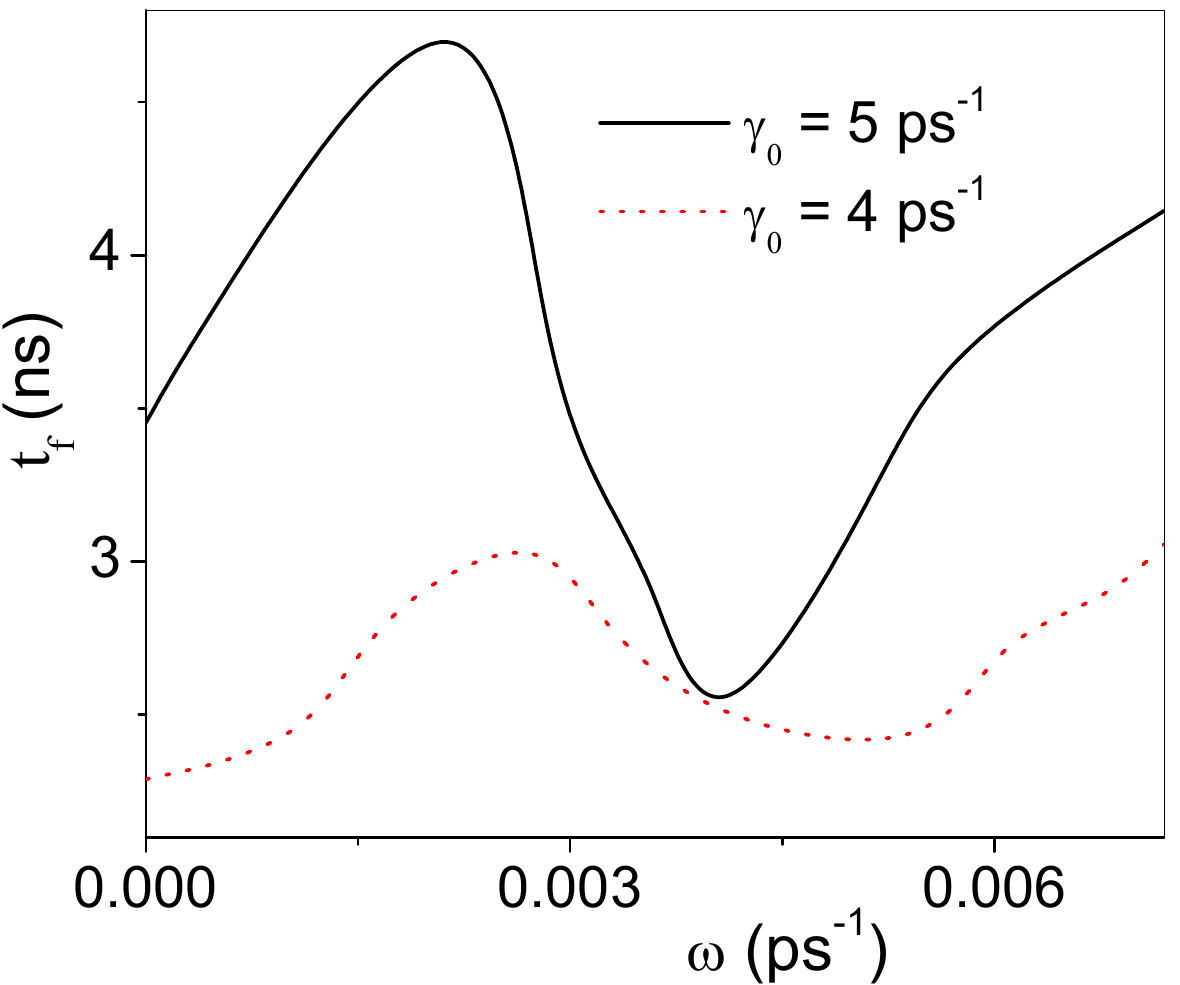,width = 8cm} \caption{(Color online) This
plot shows the dependence of mean folding time of $\beta-$ hairpin
on the frequency of the external derive in the heavy damping
situation. The chosen parameters are $T=298 K,\; \tau=1.0 \;{\rm
ps} ,\; A_0=1$ pN.} \label{fig3}
\end{figure}

\begin{figure}
\epsfig{file=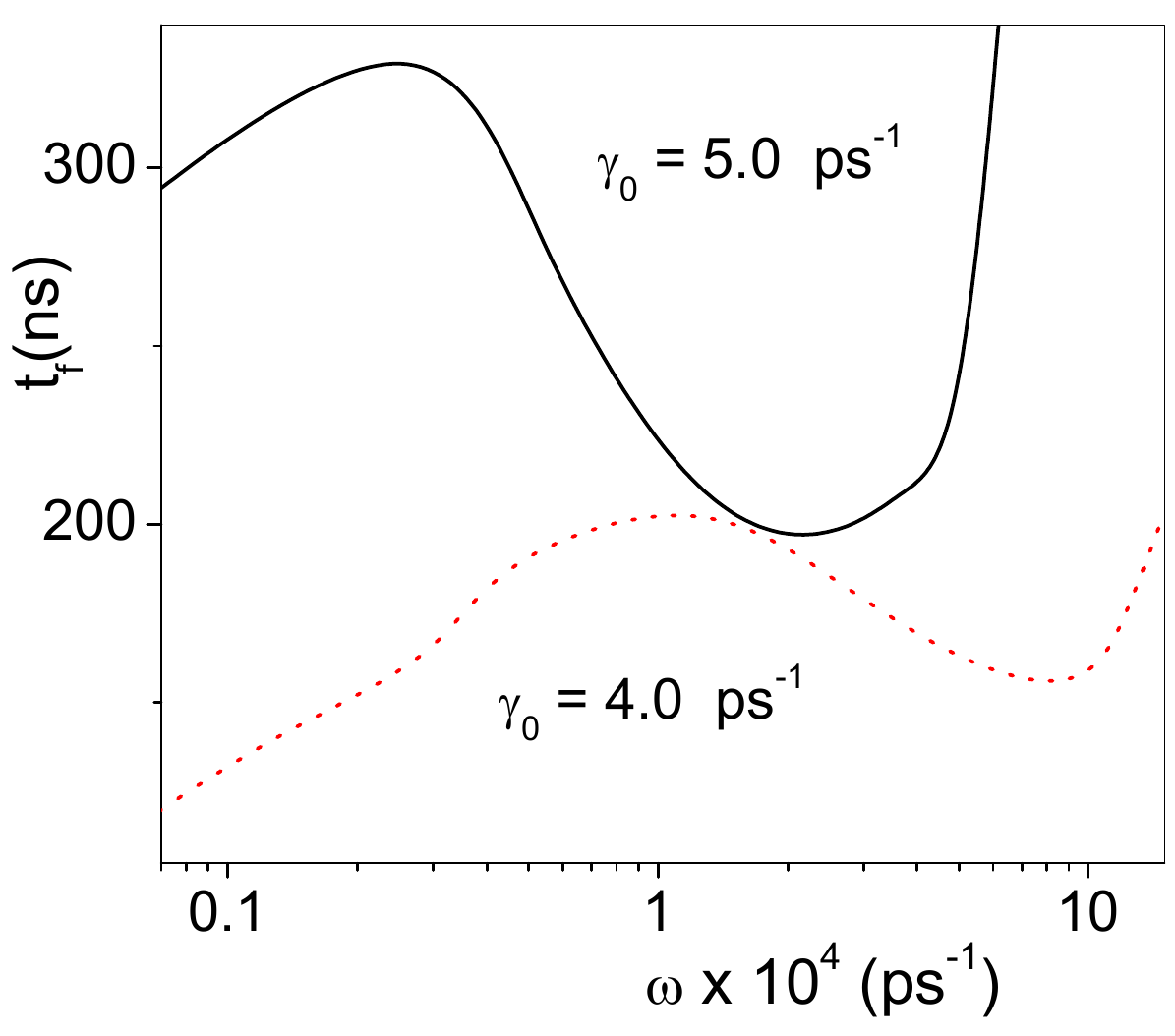,width = 8cm} \caption{(Color online) Plot of
mean folding time $t_f$ of ubiquitin vs the frequency of the
external driving force. We chose $T =298K,\; A_0=0.1$
pN,$\;\tau=1.0$ ps.} \label{fig4}
\end{figure}

Next, we have explored effects of the non-Markovian bath in the
folding dynamics in presence of external field. The folding time
of $\beta -$hairpin shows two minima with noise correlation time
(shown in Fig.~6). The reason for appearance of unusual two minima
has have discussed in detail in our recent study \cite{bagpccp}.
However, the second minimum appears at higher correlation time in
presence of the periodic force. To explain this we recall the
fluctuation dissipation relation (\ref{1.3}). It implies that the
variance of noise decreases with increase of noise correlation
time. Increase of noise correlation time leads to decrease of
fluctuations in entropy. Because of higher entropy in presence of
the driving force, the noise correlation starts playing role in
the dynamics at its larger value. Thus, the shifting of positions of the minima
are attributed to the excess entropy due to the external field.

\begin{figure}
\epsfig{file=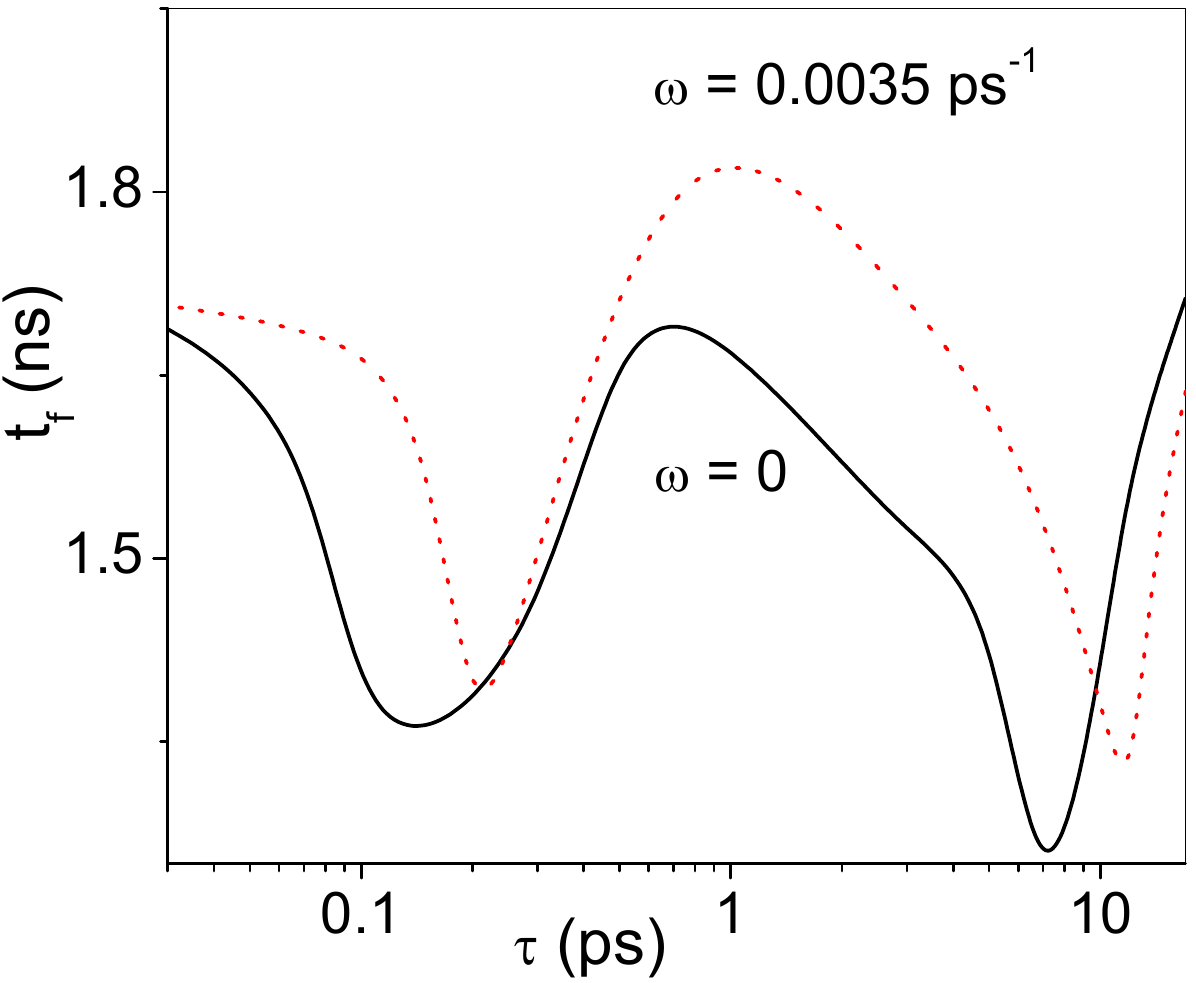,width = 8cm} \caption{(Color online) This
plot shows the dependence of mean folding time of $\beta-$ hairpin
on the correlation time of the non-Markovian noise. The chosen
parameters are $T=298 K,\; \gamma=1.0  {\rm ps}^{-1},\; A_0=1.0$
pN.} \label{fig5}
\end{figure}

 Figure 7 depicts the variation
of folding time of $\beta-$hairpin as a function of viscosity of
the medium in presence of external periodic force. We observe
that the folding time differs significantly around only the region
where the folding kinetics is accelerated due to the applied
periodic field.

\begin{figure}
\epsfig{file=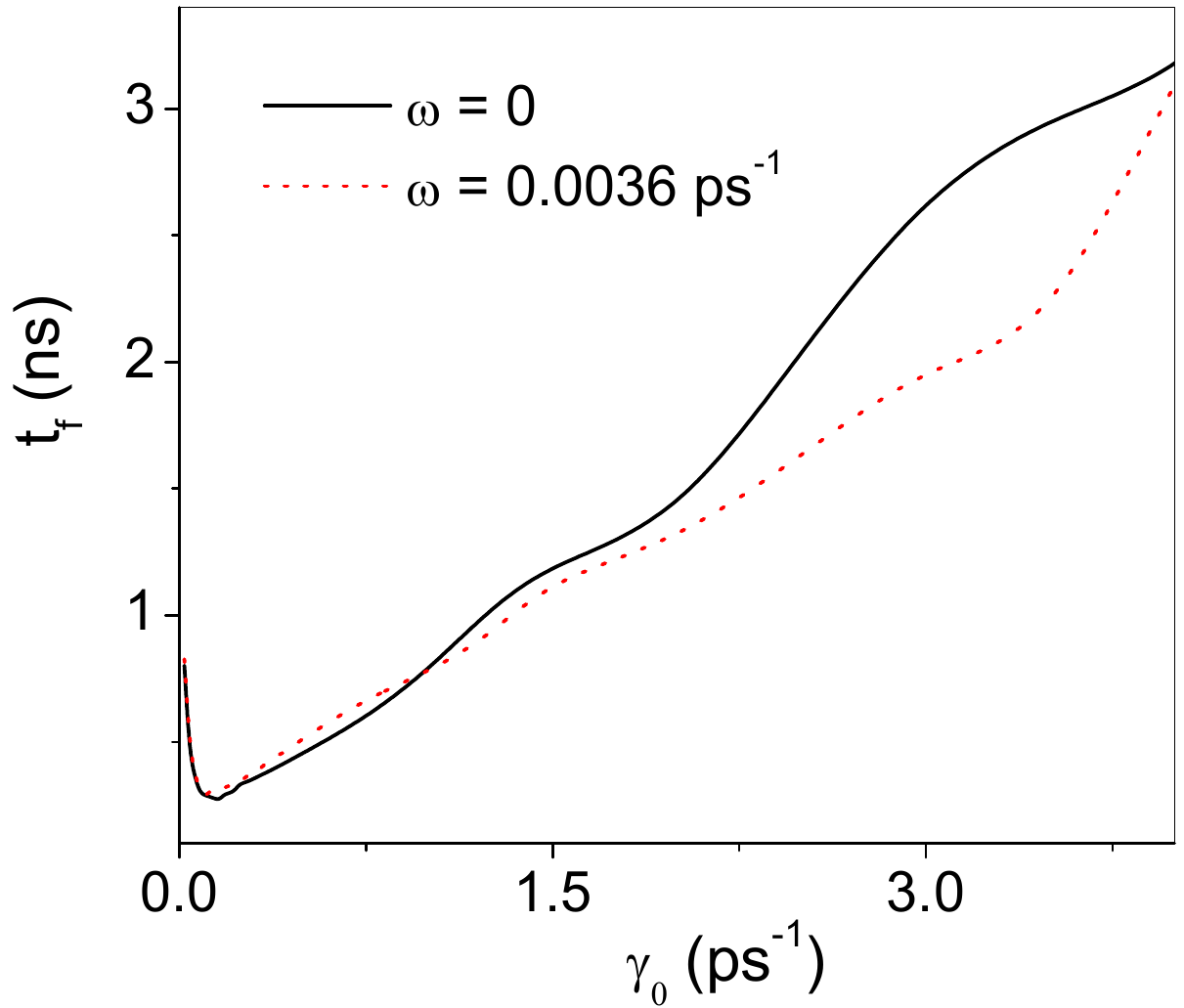,width = 8cm} \caption{(Color online) This
plot shows the dependence of mean folding time of $\beta-$ hairpin
on the viscosity of the medium. The chosen parameters are $T=298
K,\; \tau=1.0 \; {\rm ps},\; A_0=1$ pN.} \label{fig6}
\end{figure}

\begin{figure}
\epsfig{file=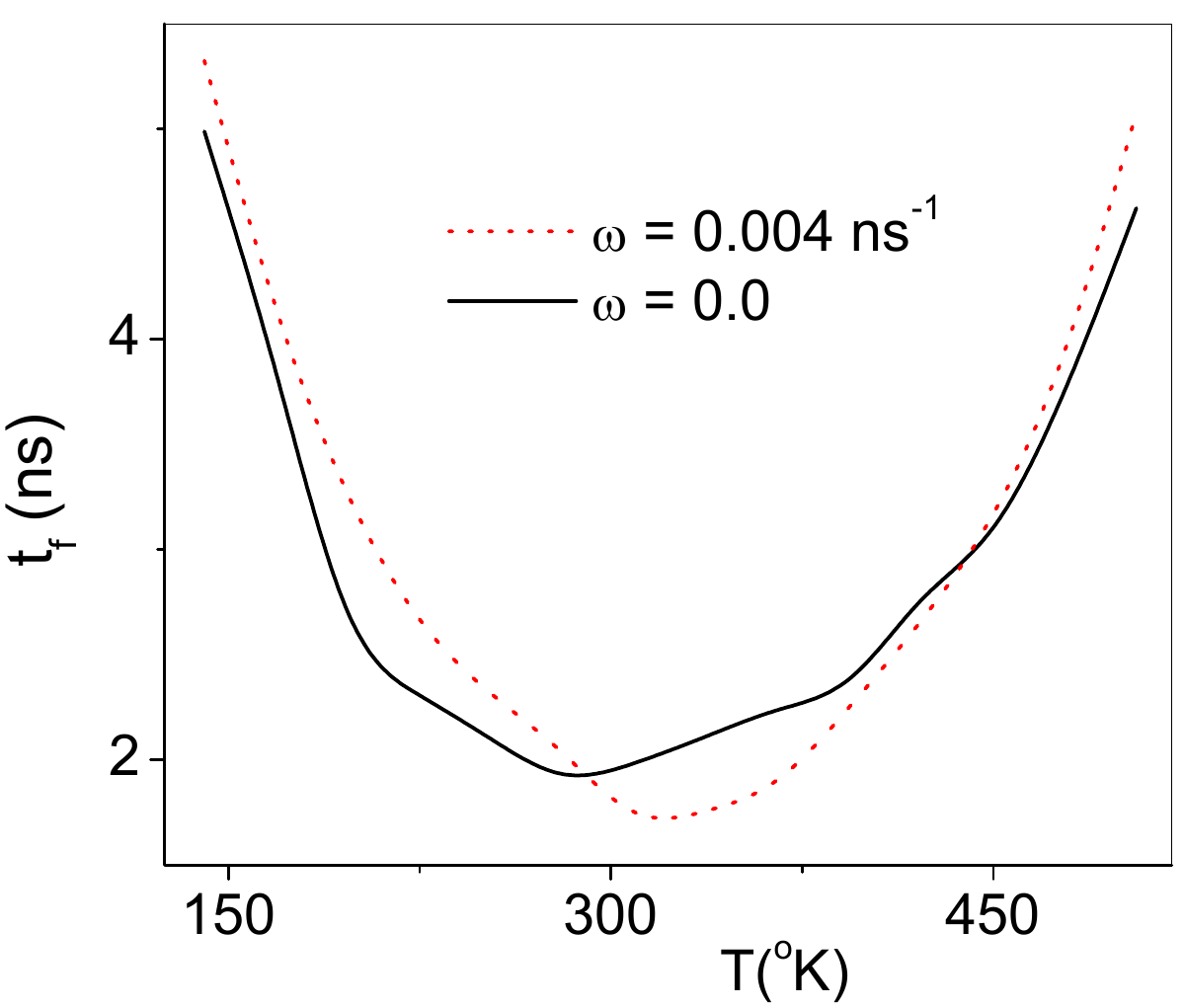,width = 8cm} \caption{(Color online) This
plot shows the dependence of mean folding time of $\beta-$ hairpin
on the Temperature of the system. The chosen parameters are
$\gamma=2
 \;{\rm ps}^{-1},\; \tau=1.0 \;{\rm ps},\; A_0=1$ pN.} \label{fig7}
\end{figure}

To elucidate the effect of external forcing on the $U$-shaped
 mean folding time versus temperature profile, we calculate the folding
time both in presence and absence of the external force. This is
depicted in Fig.~8. It shows that the folding time is greater than
the corresponding unperturbed case in the regime where increase of
entropy dominates over the energy transfer to activate towards the
transition state and is lower when the latter dominates over the
former. Acceleration of folding kinetics due to the external field
causes a shift of the minimum (see Fig.~8) towards the higher
temperature and also lowers the width of U-shaped temperature
profiles.

\section{Conclusion}
Based on the coarse-grained off-lattice models for a polypeptide
chain and setting Go-like Hamiltonian for the system we have
followed the dynamics of the protein chain by solving $N-$coupled
Langevin equations in presence of an external field. In order to
make the model more realistic we have assumed the non-Markovian
heat bath. Our main conclusions of this study are as follows:

\noindent (A) Under influence of an oscillating electric filed the
mean folding time of $\beta-$hairpin shows a maximum and a minimum
in mean folding time vs driving frequency plots. This feature is
in a sharp contrast to the one-dimensional barrier crossing
problem where only a minimum is observed in the same profile.
Thus, periodic force can induce stabilization or destabilization
of the denatured state of a protein. The folding kinetics of a
longer protein, ubiquitin, also exhibits a similar feature.
Therefore, the above observation is true for both long and small
proteins.

(B) Even in presence of an oscillating force field of amplitude
$0.1$ - $1$ pN the following three important features of
protein folding kinetics remain intact: (i) turnover behavior of
mean folding time with solvent viscosity, (ii) U-shaped mean
folding time \emph{vs.} temperature profiles, and (iii) double
minimum in mean folding time as a function of the correction of
thermal fluctuations. It implies that the above properties is very
robust to external perturbations. But the following new features
are noticed for the presence of the external force field.

\noindent (a) In the presence of the periodic force the second
minimum  appears at relatively higher noise correlation time of
the thermal noise in the plot $t_f$ {it vs.} $\tau$. This is due
to excess entropy of the system, which is produced by external
fluctuations.

\noindent (b) Increase in the damping strength enhances
acceleration of folding kinetics by the periodic force at the cost
of decreasing its robustness.

\noindent (c) Finally, we observe acceleration of the folding
kinetic due to presence of the external field in the variation of
mean folding time with viscosity of the medium and temperature.

We hope our theoretical findings could be verified experimentally
measuring protein folding time in the presence of a monochromatic
isotropic electromagnetic radiation. For a given intensity of the
radiation amino acid residues experience a resultant amplitude
vector $\vec{A}_0$. In presence of an isotropic electric field the
magnitude of $\vec{A}_0$ would be independent on the orientation
of the protein chain and time. We have considered this aspect in
our present study.



\end{document}